\newcommand{\be}{\begin{equation}}
\newcommand{\ee}{\end{equation}}
\newcommand{\ba}{\begin{eqnarray}}
\newcommand{\ea}{\end{eqnarray}}
\newcommand{\bear}{\begin{eqnarray*}}
\newcommand{\eear}{\end{eqnarray*}}
\newcommand{\lb}{\label}

\documentclass[pre,showpacs,twocolumn,superscriptaddress,showkeys]{revtex4}
\usepackage{epsfig,graphicx,psfrag,color}

\begin{document}

\title
{Two-component abelian sandpile models}
\author{F. C. \surname{Alcaraz}}
\affiliation{ Instituto de F\'{\i}sica de S\~ao Carlos, Universidade de S\~ao
Paulo, \\
Caixa Postal 369, 13560-590, S\~ao Carlos, S\~ao Paulo, Brazil}
\email{alcaraz@if.sc.usp.br}
\affiliation{Instituto de F\'{\i}sica Te\'orica,
UAM-CSIC, Madrid, Spain}
\author{P. \surname{Pyatov}}
\affiliation{Bogoliubov Laboratory of Theoretical Physics, JINR
141980 Dubna,\\ Moscow Region, Russia}
\email{pyatov@theor.jinr.ru}
\author{V.  \surname{Rittenberg} }
\affiliation{Physikalisches Institut, Bonn University, 53115 Bonn, Germany}
\email{vladimir@th.physik.uni-bonn.de}
\affiliation{Instituto de F\'{\i}sica Te\'orica,
UAM-CSIC, Madrid, Spain}
\date{\today}

\begin{abstract}
In one-component abelian sandpile models, the toppling probabilities are
independent quantities. This is not the case in multi-component models. The
condition of associativity of the underlying abelian algebras impose
nonlinear relations among the toppling probabilities. These relations are
derived for the case of two-component quadratic abelian algebras. 
We show that abelian sandpile models with two conservation laws have only trivial avalanches.

\end{abstract}

\maketitle
\section{ Introduction} \label{sect1}

Sandpile models are important toy models to understand self-organized
criticality \cite{BTW}.
In an abelian sandpile model, the toppling rules can be encoded in an
abelian algebra. This was pointed out by Dhar \cite{D1,D2,D3}. 
The structure of
these algebras \cite{AR} (see also \cite{PRU}) 
is very simple, their physical relevance will be shown 
later in the text. They are defined by taking graphs
(see Fig.~1)
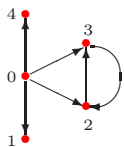
\begin{figure}[b]
\begin{picture}(200,60)(-42,0)
\put(50,32){\vector(0,1){20}}
\put(50,28){\vector(0,-1){20}}
\put(52,31){\vector(2,1){20}}
\put(52,29){\vector(2,-1){20}}
\put(73,20){\vector(0,1){21}}
\put(78,18.7){\vector(-1,0){4}}
\put(74.5,30){\oval(23,23)[r]}
\multiput(50,6.5)(0,23.7){3}{\color{red}\circle*{3}}
\multiput(73,19)(0,23.7){2}{\color{red}\circle*{3}}
\put(43,28){$\scriptstyle 0$}
\put(43,4){$\scriptstyle 1$}
\put(43,52){$\scriptstyle 4$}
\put(72,10){$\scriptstyle 2$}
\put(72,46){$\scriptstyle 3$}
\end{picture}
\caption{A typical graph on which one can define abelian algebras useful for
sandpile models. There are 4 oriented links leaving the vertex 0. The sites 2
and 3 are connected by two edges of opposite orientation.}
\end{figure}
and attaching to each vertex a generator $a_i$ of the
algebra. All the generators commute with each other. Two vertices are
connected by at most two links oriented in opposite directions. To each vertex "$i$" we
attribute a polynomial relation which expresses a power of $a_i$ (say $n$) as a polynomial in
the generators attached to the sites reached by the
outgoing arrows starting at  "$i$"
as well as $a_i$.
 The degree of the polynomial is at
most equal to $n$. This implies for example that for the vertex "0" in Fig.~1
we have
\be
\lb{1}
a_0^n\, =\, P(a_0,a_1,a_2,a_3,a_4). 
\ee
In the corresponding sandpile
model, $a_0^k$ is interpreted as having $k$ grains of sand on the vertex $0$.
The coefficients in the polynomial $P(a_0, a_1, a_2, a_3, a_4)$ are
nonnegative and their sum is equal to 1, as we are going to see, 
 they are going to be interpreted as probabilities.

A simple and very relevant example \cite{BTW} is the case of a two-dimensional
square lattice (coordinates $(i,j)$) the abelian algebra being given by the
relations:
\be
\lb{2}
a_{i,j}^4 = a_{i+1,j}\,a_{i,j+1}\,a_{i-1,j}\,a_{i,j-1} ,\quad
\left[a_{i,j},\,a_{i',j'}\right] = 0 .
\ee
We didn't specify the boundary conditions.

The sandpile model is defined by a stochastic
process which gives the stationary state and the rules how the
sand grains act.

In continuous time and a lattice with $N$ vertices (sites) the time
evolution of the system is given by a Hamiltonian $H$:
\be
\lb{3}
H\, =\, \sum_{i=1}^N w_i (1-a_i)\, ,
\ee
where the nonnegative coefficients $w_i$ 
are transition rates and we chose the unit of time by taking:
\be
\lb{4}
\sum_{i=1}^N w_i \, =\, 1\, .
\ee
The Hamiltonian (\ref{3}) acts in an $M$-dimensional vector space given by all
the independent $M$ monomials in the generators $a_i$. 
There is a correspondence between each
monomial and 
a configuration in the sandpile model in which to each generator 
appearing
in the monomial corresponds sand grains on the respective sites. Their
number equals the power of which the generator appears in the monomial

 The action of $H$ on the configuration space can be understood in the
following way: take a configuration, in a unit of time, with a probability $w_i$ a grain of sand 
is added at the site $"i"$. As a result this configuration goes to others,
with probabilities given by the toppling rules (\ref{1}). 

The unnormalized probabilities $P_m(t)$ to find the system in a configuration "$m$" at the
time $t$ can be obtained from the master equation:
\be
\lb{5}
\frac{dP(t)}{dt}\, =\, - H P(t)\, ,
\ee
where
\be
\lb{6}
P(t)\, =\, \sum_{m=1}^M P_m(t) W(m)\, ,
\ee
$W(m)$ is the monomial corresponding to the configuration "$m$".

The stationary PDF will be denoted by $|0\rangle$ ($H|0\rangle = 0$).

Except for positivity there are no supplementary constraints in the
relations (\ref{1}). This implies that one can choose at will the toppling
probabilities, however as we are going to see this is {\em not}  the case if one
considers two-component sandpiles models.

\section{Sandpile models with two kinds of sand}

In  two-component abelian sandpile models, one assumes that one has two types of sand
(say "$a$" and "$b$"). One adds with a probability $u_i$ ($v_i$) a grain
of sand "$a$" (respectively "$b$") to the site "$i$". 
Grains topple if there are more than one, of any kind, on a given  site. In the toppling process,
the two types of sand mix and eventually transform into each other.
 We are going to consider the simplest model of
this kind.

Consider a one-dimensional directed lattice (Fig.2)
\begin{figure}[t]
\begin{picture}(200,40)
\multiput(50,10)(23,0){2}{\vector(1,0){20}}
\multiput(49,10)(23,0){3}{\color{red}\circle*{3}}
\multiput(109,10)(6,0){3}{\circle*{1}}
\multiput(135,10)(23,0){2}{\color{red}\circle*{3}}
\put(136,10){\vector(1,0){20}}
\put(47.5,0){$\scriptstyle 1$}
\put(70.5,0){$\scriptstyle 2$}
\put(93.5,0){$\scriptstyle 3$}
\put(126.5,0){$\scriptstyle N-1$}
\put(156.5,0){$\scriptstyle N$}
\end{picture}
\caption{A one-dimensional acyclic directed lattice with $N$ sites. All edges
have the same orientation.}
\end{figure}
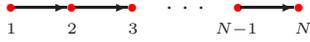
with $N$ sites \cite{AR}. To each
site "$i$" we attach two generators $a_i$ and $b_i$, all of them being mutually commuting
\be
\lb{7a}
[a_i, a_{i'}]\, =\, [a_i, b_{i'}]\, =\, [b_i, b_{i'}]\, =\, 0\,\; \forall\, i,i'=1,\dots ,N.
\ee
For simplicity we consider quadratic algebras only.
The most general quadratic relations involving nearest-neighbor
interactions only are
\begin{equation}
\lb{7}
\begin{array}{rcl}
a_i^2 &=& \hspace{-.1mm}\alpha_1\, a_i a_{i+1}\, +\, \alpha_2\, a_i b_{i+1}\, +\,\beta_1\, b_i a_{i+1}\, +\,
\beta_2\, b_i b_{i+1}\\[.5mm] &&\hspace{15.3mm}
+\, \xi_1\, a_{i+1}^2 \, +\, \xi_2\, b_{i+1}^2\, +\, \xi_3\, a_{i+1}b_{i+1},
\\[2mm]
b_i^2 &=&\hspace{.3mm} \gamma_1\, a_i a_{i+1}\, +\,\gamma_2\, a_i b_{i+1}\, +\,
\delta_1\, b_i a_{i+1}\, +\,\delta_2\, b_i b_{i+1}\\[.5mm] &&\hspace{15.3mm}
+\, \eta_1 a_{i+1}^2\, +\, \eta_2\,b_{i+1}^2\, +\,\eta_3 \,a_{i+1}b_{i+1},
\\[2mm]
a_i b_i &=& \mu_1\, a_i a_{i+1}\, +\, \mu_2\, a_i b_{i+1}\, +\,
\nu_1\, b_i a_{i+1}\, +\, \nu_2\, b_i b_{i+1}\\[.5mm] &&\hspace{15.3mm}
+\,\zeta_1 a_{i+1}^2\,  +\,\zeta_2\,b_{i+1}^2\, +\, \zeta_3 \,a_{i+1}b_{i+1},
\end{array}
\end{equation}
for $i=1,\ldots,N$, 
 in which we take:
\be
\lb{8}
a_{N+1}\, =\, b_{N+1}\, =\, 1\, .
\ee
This implies that the two types of sand may leave the
system on the site $N$. 
The constants in (\ref{7}) are positive probabilities, so that
\be
\nonumber
\alpha+\beta+\xi\,=\,\gamma+\delta+\eta\,=\,\mu+\nu+\zeta\, =\, 1\, ,
\ee
where we used the notation $\alpha=\sum_{i=1}^2\alpha_i$, $\beta=\sum_{i=1}^2\beta_i$,
$\xi=\sum_{i=1}^3\xi_i$, etc.

The relations (\ref{7a})--(\ref{8}) don't define yet an algebra since one has
still to impose associativity (diamond conditions):
\be
\lb{9}
(a_i^2)b_i\, =\, a_i(a_ib_i)\, ,\quad
a_i(b_i^2)\, =\, (a_ib_i)b_i\, .
\ee
Introducing (\ref{7}) in (\ref{9}) we find the following 12 relations:
\begin{equation}
\lb{10}
\begin{array}{rcl}
\xi_i &=& \beta_i(\mu_i-\delta_i)+\nu_i(\nu_i-\alpha_i)\, ,\quad i=1,2,
\\[1mm]
\xi_3 &=& \beta_1(\mu_2-\delta_2)+\beta_2(\mu_1-\delta_1)\\
&&\hspace{18mm}+\nu_1(\nu_2-\alpha_2)+\nu_2(\nu_1-\alpha_1)\, ,
\\[2mm]
\eta_i &=& \gamma_i(\nu_i-\alpha_i)+\mu_i(\mu_i-\delta_i)\, ,\quad i=1,2,
\\[1mm]
\eta_3 &=& \gamma_1(\nu_2-\alpha_2)+\gamma_2(\nu_1-\alpha_1)\\
&&\hspace{18mm}+\mu_1(\mu_2-\delta_2)+\mu_2(\mu_1-\delta_1)\, ,
\\[2mm]
\zeta_i &=& \beta_i\gamma_i-\mu_i\nu_i\, ,\quad i=1,2,
\\[1mm]
\zeta_3 &=& \beta_1\gamma_2+\beta_2\gamma_1-\mu_1\nu_2-\mu_2\nu_1\, ,
\end{array}
\end{equation}
and
\ba
\lb{11a}
\beta(1+\mu-\delta)&=&(1-\nu)(1+\nu-\alpha)\, ,
\\
\lb{11b}
\gamma(1+\nu-\alpha)&=&(1-\mu)(1+\mu-\delta)\, ,
\\
\lb{11c}
\beta\gamma&=&(1-\mu)(1-\nu)\, .
\ea
Out of the 3 relations (\ref{11a})--(\ref{11c}) only two are independent since multiplying
(\ref{11a}) with (\ref{11b}) one obtains (\ref{11c}).

As one can see from (\ref{10})--(\ref{11c}) in the two-component case the various
toppling probabilities are constrained by nonlinear relations.

The vector space in which the Hamiltonian
\be
\lb{12}
H\, =\, \frac{1}{N}\sum_{i=1}^N\, (1\,-\,u_i\,a_i\, -\, v_i\,b_i)\, ,\quad
u_i+v_i=1,
\ee
acts is made out of monomials in which either $a_i$, $b_i$ or 1 appears
for a given site "$i$". The physical interpretation of these monomials is
obvious. If in the monomial $a_i$ ($b_i$) appears, on the site "$i$" one has a
grain of sand of type "$a$" ("$b$"). If neither $a_i$ or $b_i$ appear in the
monomial, there is a vacancy on the site "$i$".

It is easy to show the the stationary state PDF is of product form:
\be
\lb{13}
{
|0\rangle  ={\textstyle
\prod_{i=1}^N
\frac{(1-\alpha)(1-\delta)-\mu\nu+
(1+\mu-\delta)a_i+(1+\nu-\alpha)b_i}{(2-\alpha)(2-\delta)-(1-\mu)(1-\nu)}}
}\, .
\ee

An obvious question is if there is a solution of (\ref{10})--(\ref{11c}) in which each
type of sand is conserved separately. This would imply perhaps a new
universality class of sandpile models. If one takes only
$\alpha_1$, $\xi_1$, $\delta_2$, $\eta_2$, $\mu_2$, $\nu_1$ and $\zeta_3$ as non vanishing
probabilities, in the bulk, the number of sand grains of type "$a$" and "$b$"
are conserved separately.
There are only two equivalent solutions of the diamond conditions in this case.
 One
solution is:
\ba
\lb{14a}
a_i^2 &=& \alpha\, a_ia_{i+1}\, +\, (1-\alpha)\, a_{i+1}^2\, ,
\\[1mm]
\lb{14b}
b_i^2 &=& b_i b_{i+1}\, ,
\\[1mm]
\lb{14c}
a_i b_i &=& b_i a_{i+1}\, .
\ea
In the second solution, one exchanges the generators $a_i$ with $b_i$.

The stationary PDF of the stochastic process (\ref{12}),
(\ref{14a})--(\ref{14c}) is
\vspace{-4mm}
\be
\lb{15}
|0\rangle\, =\, \prod_{i=1}^N b_i\, .
\ee
There are only trivial  avalanches in this case. If a grain of sand of type "$a$" hits the
system, it leaves directly at the boundary because of (\ref{14c}), 
leaving the system unchanged.
If a sand grain of type "$b$" hits the system, it also leaves it directly
at the boundary because of (\ref{14b}).
This is a surprising result.

We consider now a more general graph where the vertex "$i$"
is linked by outgoing arrows to a number of sites which
we label by index "$x$". If
one asks for two conservation laws and imposes
the associativity condition (\ref{9}),
the generalization of the solution (\ref{14a})--(\ref{14c}) reads:
\ba
\lb{14a-gen}
a_i^2 &=& \sum_{x}\alpha_x\, a_ia_{x}\, +\, \sum_{x,y}\xi_{x,y}\, a_x a_y\, ,
\\
\lb{14b-gen}
b_i^2 &=& \sum_x \delta_x\,b_i b_{x}\, ,
\\
\lb{14c-gen}
a_i b_i &=& \sum_x\nu_x\, b_i a_{x}\, ,
\ea
where the toppling probabilities satisfy the relations
$\sum_x \delta_x\, =\, \sum_x\nu_x\, =\, 1 ,\;$ $\alpha_x\leq \nu_x\, ,\;$ and
\be
\lb{14-add}
\xi_{x,y}\, =\, {1\over 2}\bigl\{ \nu_x(\nu_y-\alpha_y) +\nu_y(\nu_x-\alpha_x)\bigr\} .
\ee
Notice that, similar to the one-dimensional case, only the grains of sand
of type "a" topple. This has as consequence that for an arbitrary
graph the stationary state of the system is an absorbing state, like (\ref{15}), in which
only particles "b" are present. The proof is straightforward. 
Only trivial 
avalanches can be obtained.

We have shown that using other representations of the algebra
(\ref{14a})--(\ref{14c}) which may define other sandpile
models (see \cite{AR}), one obtains the same result: two conservation laws
are compatible with trivial  avalanches only. The system stays unchanged in the aftermath.

We have checked, only for quadratic algebras, that for a vertex connected
by arrows to several vertices (there was only one in example (\ref{7})), the
associativity conditions (\ref{9}) are incompatible with the existence of  nontrivial avalanches in the
sandpiles with two conservation laws. This is a surprising result. We have
no proof of a similar statement for more general algebras (cubic, quartic,\ldots ).
Simple examples like the generalization of (\ref{2}), give the same result.

If we  don't insist on two conservation laws, there are various solutions
of the constraints (\ref{10})--(\ref{11c}) and therefore  two-component sandpile
models. They all share the property that during the toppling process,
grains of sand of type "$a$" and "$b$" mutate
(we have not identified a physical process which leads to such a
phenomenon).
 We present a simple example in
which we consider the abelian algebra:
\ba
\nonumber
a_i^2 &=& (1-\nu)\, a_i a_{i+1}\, +\, {\nu\over 1+\mu}\, b_i a_{i+1}
\\[0mm]
\lb{16a}
&&\hspace{3mm}
+\, {\nu\phi\over 1+\mu}\, a_{i+1}^2 \,  +\, {\nu(\mu-\phi)\over 1+\mu}\, a_{i+1}b_{i+1},
\\[1mm]
\nonumber
b_i^2 &=&(1-\mu)\,  b_i b_{i+1}\,
+\,\phi(1+\mu-2\phi) \,a_{i+1}b_{i+1}
\\[1mm]
\lb{16b}
&&\hspace{3mm}
+\, \phi^2 a_{i+1}^2\, +\, (1-\phi)(\mu-\phi)\,b_{i+1}^2,
\\[2mm]
\lb{16c}
a_i b_i &=& \phi\, a_i a_{i+1}\, +\, (1-\phi)\, a_i b_{i+1}\, .
\ea
and the Hamiltonian (\ref{12}) with $u_i = v_i = 1/2$.
This implies that we take the stationary state
\be
\lb{18}
|0\rangle = \prod_{i=1}^N{\mu\nu + (1+\mu)a_i + \nu b_i\over (1 +\mu)(1+\nu)}
\ee
and add with equal probability a grain of sand of type "$a$" and "$b$" on the
first site. Notice that $|0\rangle$ is independent on $\phi$.

We are interested to know what is the probability $P_a(T)$ to have an
avalanche ending with a grain "$a$" and having a 
 duration $T$. The duration of the avalanche, in this one dimensional case, is the number of sites where the topplings occur.
  $P_b(T)$ has a similar
meaning. In \cite{AR} it was shown that in the case of the one-component model
which is obtained by considering only (\ref{16b}) in which one takes $\phi = 0$,
at large values of $T$, one has
\be
\lb{19}
P_b(T) \sim T^{-3/2}\, .
\ee
This means that for the one-component case the avalanches are in the
random walker universality  class \cite{MZ}.
\begin{figure}[t]
\begin{center}
\begin{picture}(250,240)
\put(0,0){\epsfxsize=240pt\epsfbox{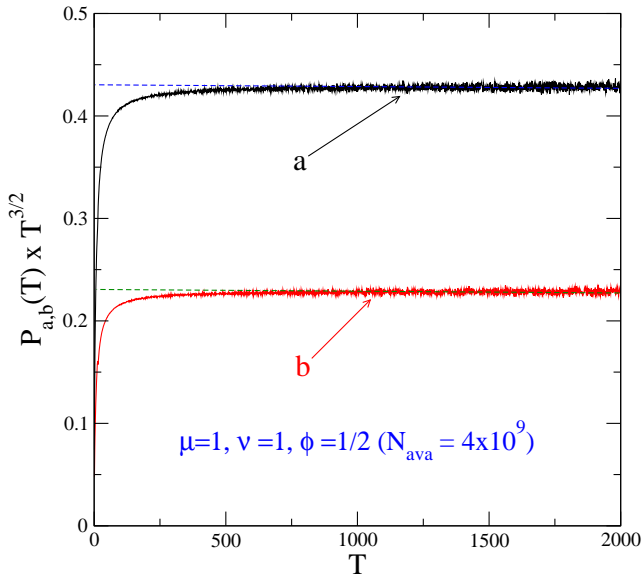}}
\end{picture}
\caption{(Color online) Probability distribution functions $P_a(T)$ and $P_b(T)$ to have
avalanches with duration  $T$ ending with grains of sand of type "$a$", respectively,
"$b$" obtained in the model given by (\ref{16a})--(\ref{16c}). The values $\mu = \nu = 1$ and
$\phi =1/2$ were used for the parameters of the model. $4\times 10^9$ avalanches were
observed in the Monte Carlo simulations. In the figure $P_a(T)$ and $P_b(T)$
are multiplied by $T^{3/2}$ in order to show their large $T$ behavior.}
\end{center}
\end{figure}
In Fig.3 we present the results of
Monte-Carlo simulations for the model given by (\ref{16a})--(\ref{16c}) in which we have taken
$\mu = \nu = 1$, $\phi =1/2$. We observe that for large values of $T$,
\be
\lb{20}
P_a(T) \sim T^{-3/2},\qquad  P_b(T) \sim T^{-3/2}\, ,
\ee
mutations between sand grains of type "$a$" and "$b$" don't change the
universality class. This stays valid for the whole parameter space except for the boundaries
$\phi = 0$ and $\phi = \mu$.

If in (\ref{16a})--(\ref{16c}) one takes $\phi = 0$, $P_a(T)$ decreases exponentially and
$P_b(T)$ has an algebraic falloff (\ref{19}). This can be easily shown using the
algebra (\ref{16a})--(\ref{16c}) and the recurrence relations for the toppling
probabilities which follow. Conversely, if $\phi=\mu$, $P_b(T)$ decays exponentially and
$P_a(T)$ has the algebraic falloff.

\section{Conclusions}

The main message of this paper is that if one is interested in
two-component abelian sandpile models, the toppling probabilities are not
arbitrary. They have to \mbox{satisfy} nonlinear relations coming from the
condition that the algebra is associative. These constraints don't exist in
the one-component models. An unexpected result is that, at least for the case of
quadratic algebras, the co-existence of two conservation laws in
the bulk (one for each component), and of nontrivial avalanches, is impossible.

We have also shown, in an example of a one-dimensional directed model,
that once we allow the two components
to mutate in each other during the toppling process, the one-component and
the two-component models belong to the same universality class.

We believe that our conclusions apply to any multi-component models.

We would like to mention a possible extension of the quadratic algebra 
(8). If we omit the last of the three equations (8), two grains of sand, 
one of type "a" and the other one of type "b", don't topple. As a result 
the stationary state is a linear combination of $4^N$ instead of $3^N$ states.
If we are not interested in mutations ($a \leftrightarrow b$), one has a direct product 
of two algebras. One contains only $a_i$ generators (like (18)) and the other 
one contains only $b_i$ generators. One gets avalanches in which grains 
of type "a" and "b"  don't mix.
If one consider the possibility of mutations the situation is 
different. Take for example the algebra defined by (26) and (27)
(we have omitted (28)!), the total number of "grains" is not conserved 
since $a_{i+1}b_{i+1}$ has to be looked upon as a new kind grain. 
As a result, the 
PDFs
of the duration of avalanches ending in a grain of sand of type "a", "b" or 
"ab" have an exponential falloff. We have also checked that if we change the 
algebra by allowing mutations, while conserving the number of sand grains, 
the avalanches belong to the random walker universality class \cite{MZ}.

\section*{Acknowledgments}
We would like to thank D. Dhar for reading the manuscript and 
relevant  comments.
The work of F.C.A. was
partially supported by FAPESP and CNPq (Brazilian Agencies).
The work of P.P and V.R. was supported by the  DFG-RFBR grant
(436 RUS 113/909/0-1(R) and 07-02-91561a)
and by the grant of the Heisenberg-Landau program. F.C.A and V.R. thank 
the kind hospitality of IFT-UAM/CSIC - Madrid - Spain, where this paper was concluded.




\begin{thebibliography}{99}
%
\bibitem{BTW} P. Bak, C. Tang and K. Wiesenfeld,
Phys. Rev. Lett. {\bf 59},381 (1987); D. Dhar and R. Ramaswamy, Phys. Rev. Lett. {\bf 63}, 1659 (1989).
%
\bibitem{D1} D. Dhar,
Physica A {\bf 263},4  (1999).
%
\bibitem{D2} D. Dhar,
Physica A {\bf 270},1 (1999), and references therein.
%
\bibitem{D3}T. Sadhu and D. Dhar, {\tt arXiv:Cond-mat/0809.261}.
%
\bibitem{AR} F. C. Alcaraz and V. Rittenberg,  Phys. Rev. E {\bf 78}, 032319 (2008).
%
\bibitem{PRU} G. Pruessner, J. Phys.A {\bf 37}, 7455 (2004); 
M. Stapleton and K. Christensen, Phys. Rev. E {\bf 72}, 66133 (2005).
%
\bibitem{MZ} S. Maslov and Y-C. Zhang,
Phys. Rev. Lett. {\bf 75},1550 (1995).
%
\end{thebibliography}
\end{document}